\documentclass[twoside,11pt]{article}
\usepackage{amsmath,amssymb,amsfonts}
\usepackage{amsfonts}
\usepackage{graphicx}
\usepackage{graphics}
\usepackage{breqn}
\usepackage{microtype}
\usepackage{setspace}
\begin{document}
	
	\setlength{\unitlength}{10mm}
	\newcommand{\f}{\frac}
	\newtheorem{theorem}{Theorem}[section]
	\newcommand{\sta}{\stackrel}

	\title{Calculation of binding energy and wave function for exotic hidden-charm pentaquark}
	
	\author{F. Chezani Sharahi$^{1, 2}$\thanks{spin1631@gmail.com}, M. Monemzadeh$^{1}$\thanks{Corresponding author: Majid Monemzadeh; monem@kashanu.ac.ir} \\
		\it\small{{$^{1}$Department of  Physics,  University of kashan, Iran}}\\
		\it\small{{$^{2}$ Department of Physics,  University of Arak, Iran}}}

	
	\date{}
	\maketitle
	\vspace{1cm}
	\vspace{-1.5cm}\begin{abstract}
		\noindent  
		In this study, pentaquark $P_{c}(4380)$ composed of a baryon $\Sigma_{c}$ and a $\overline{D}^{*}$ meson is considered. Pentaquark is as a bound state of two-body systems composed of a baryon and a meson. The calculated potential will be expanded and replaced in the Schr\"{o}dinger equation until tenth sentences of expansion. Solving the Schr\"{o}dinger equation with the expanded potential of Pentaquark leads to an analytically complete approach. As a consequence, the binding energy $E_{B}$ of pentaquark $P_{c}$ and wave function are obtained. The results $E_{B}$ will presented in the form of tables so that we can review the existence of pentaquark $P_{c}$. Then, the wave function will be shown on diagrams. Finally, the calculated results are compared with another obtained results and the mass of observing pentaquark $P_{c}$ and the radius of pentaquark are estimated.	
		

	\end{abstract}


\newpage

\section{Introduction}
 
 In the first researches, the existence of multi-quark states are illustrated in its simplest possible form, in which baryons made of three fundamental quarks and mesons from a quark and an antiquark\cite{1}. Indeed, from the point of view of the mathematics and physics, there was no QCD theorem opposing the existence of exotic multi-quark. In gauge field theory, the QCD principle allows the existence of multi - quarks and hybrids, which include quark and gluonic degrees of freedom\cite{2}. \\
 
Searching to find the pentaquark and its probing has a long history. About ten years ago, in the study of pentaquark, a big progress was occurred when LEPS collaborations getting started with a claim to find out strong evidence of the pentaquark with the mass of about $1.540 GeV$\cite{3}. Then, many theoretical and experimental approaches were pursued and many ideas were proposed in this field. For example, Zou and his colleagues suggested that the components of pentaquark be included with nucleon. Since the heavy quarks play an important role in stabilization of multi - quark systems and these play exactly the same role as the hydrogen molecule in QED\cite{4, 5, 6, 7}, so there are theoretical predictions about exotic hidden-charm pentaquark. In particular, the possibility existence of hidden-charm molecular baryons composed of an anti-charmed meson and a charmed baryon was systematically studied within the one boson exchange model\cite{3}. \\ 

In 2003, LEPS collaboration reported the evidence of pentaquark $s$ state with content of quark $uudd\bar{s}$ and very narrow width\cite{8}. Unfortunately, this exotic flavor was not confirmed in subsequent experiments\cite{9, 10}. In fact, the possible theoretical arguments are presented for the non-existence of a stable pentaquark $s$ in the references\cite{11, 12}. Also, this mode has not been found with a light flavor yet. However, the baryons with light flavor may be able to have a significant pentaquark components\cite{2}.\\

One decade ago, numerous researches were done in around of the world to find exotic particles. The result of these efforts was observation the mesonic X , Y , Z particles in  Belle, BESIII, BABAR and LHCb. Some of these were considered as candidates for exotic states because they don't fit in a regular mesonic structure\cite{1}. The common point of the exotic states is that all of these contain the heavy quarks and antiquarks.\\

Due to the heavy quarks, the exotic states can be stable and light modes can be combined with a regular mode\cite{13}. This guess is consistent with the fact that all of the exotic states have a  hidden  $c$ or $b$ which this is experimentally observed. If this claim is valid that the heavy components stabilize multi-quark systems by particle physics scientists, it will a natural proposal from their experimental colleagues to be search for exotic states consist of clear quarks $c$ and $b$, for example $b \bar{c}qq'$ and etc\cite{2}.\\

\begin{figure}
	
	\includegraphics[width=12 cm, totalheight=4.2 cm]{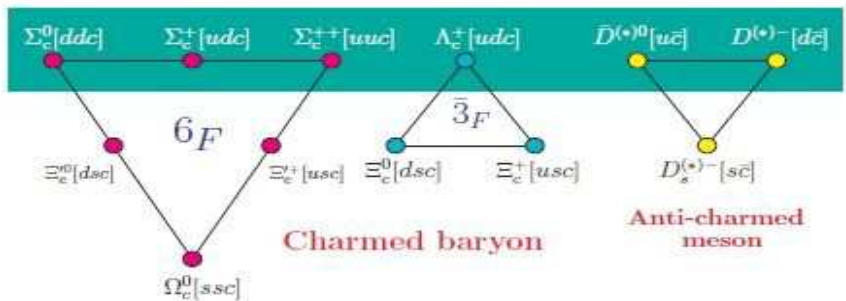}
	\caption{\it\small{{(Color line)The s-wave charmed baryons with  $J^{p}=1/2^{+}$ and  pseudo- scalar or vector s-wave charmed mesons which form the molecular baryons with double-charm\cite{6}.}}}
	\label{fig:1}
\end{figure}

Here, one hidden-charm molecular baryon composed of an anti-charmed meson and a charmed baryon has been studied. This can have one of the two flavors i.e. symmetric $6_{F}$ or anti-symmetric flavor $\bar{3}_{F}$ as shown in Fig. 1. So spin-parity of s-wave charmed baryon is $J^{p}=1/2^{+}$ or $J^{p}=3/2^{+}$ for $6_{F}$ and $J^{p}=1/2^{+}$ for $\bar{3}_{F}$. The pseudo- scalar or vector anti-charmed meson is made of s-wave anti-charmed mesons. In Figure 1, hidden-charm molecular states composed of the anti-charmed mesons and the charmed baryons are placed inside the green range\cite{6}.\\

Recently, LHCb collaboration has observed two resonance structures  $P_{c}(4380)$ and $P_{c}(4450)$ with mass and width decay $M_{P_{c}(4380)}=4380\pm 8\pm 29 MeV$, $\Gamma_{P_{c}(4380)}=205\pm 18\pm 86 MeV$ and $M_{P_{c}(4450)}=4449.8\pm 1.7\pm 2.5 MeV$, $\Gamma_{P_{c}(4450)}=39\pm 5\pm 19 MeV$ in the invariant mass spectrum $J/\psi p$ from $\Lambda_{b}\rightarrow J/\psi p K$. According to the final state $J/\psi p$ is concluded that two observing states $P_{c}$, unable to be isosinglet and these are consist of hidden-charm quantum numbers. Also, it is suggested structures for each of the states $P_{c}$ which is considered for $P_{c}(4380)$, $\Sigma_{c}(2455)\overline{D}^{*}$ and $P_{c}(4450)$, $\Sigma_{c}^{*}(2520)\overline{D}^{*}$\cite{2, 3}. \\

 Binding energy may be calculated analytically by solving Schr\"{o}dinger equation with the expanded potential of Pentaquark. This approach offers advantages over numerical solution of the Schr\"{o}dinger equation for pentaquark. First all values for binding energy can be calculated and this is done more accurately. Second wave function can be computed and presented graphically ($\phi (r)$ versus $r$), from which useful particle characteristics and data could be extracted.\\
 	
In this work, it is described by analytical solution of pentaquark $\Sigma_{c}\overline{D}^{*}$ in four sections: where apart from the introduction as section 1, the potential of pentaquark $P_{c}(4380)$ is discussed in section 2. Then, the Schr\"{o}dinger equation using the potential of pentaquark is analytically solved in section 3. Finally, important conclusions are discussed in section 4.\\

\section{ The potential of pentaquark $P_{c}$ and its expansion}

Pentaquark is considered to be composed of one baryon and one meson. For pentaquark, with such a structure, the potential is given as follows\cite{2}:  

\begin{equation}
V_{\Sigma_{c}\overline{D}^{*}}(r)= \frac{1}{3}\ \frac{g g_{1}}{f_{\pi}}\ \nabla^{2}Y(\Lambda, m_{\pi}, r)\mathcal{J}_{0} \mathcal{G}_{0},      
\end{equation}
 
   where, the coupling constant $g=0.59 \pm 0.07 \pm 0.01$ is extracted from the width of $D^{*}$ \cite{14, 15}, $g_{1}= 0.75, 0.94, 1.95$ \cite{6, 16}. Also, it is the mass of pion $m_{\pi}=135 MeV$ and pion decay constant $f_{\pi}=132 MeV$ \cite{2}. The amount of phenomenological cutoff parameter is considered $\Lambda=0.8 GeV-2.5 GeV$ \cite{2, 12}. Finally, the $Y(\Lambda, m_{\pi}, r)$ is \cite{2}: 

\begin{eqnarray}
Y(\Lambda, m_{\pi}, r)= \frac{1}{4 \pi r}\ (e^{-m_{\pi}r} - e^{-\Lambda r})-\frac{\Lambda^{2}-m_{\pi}^{2}}{8 \pi \Lambda}\  e^{-\Lambda r}.
\end{eqnarray}

Now, we are calculating and expanding $\nabla^{2}Y$:

\begin{eqnarray}
\nabla^{2}Y(\Lambda, m_{\pi}, r)&=& \frac{m_{\pi}^{2}}{4 \pi r}\ (e^{-m_{\pi}r} - e^{-\Lambda r})-\frac{\Lambda^{3}-m_{\pi}^{2}\Lambda}{8 \pi}\  e^{-\Lambda r}\nonumber\\
&=&\frac{1}{4 \pi}(b_{0}+b_{1}r+b_{2} r^{2}+b_{3} r^{3}+b_{4} r^{4}+ b_{5} r^{5}\nonumber\\
&&+b_{6} r^{6}+b_{7} r^{7}+b_{8} r^{8}+b_{9} r^{9}+...).
\end{eqnarray} 

Where:

\begin{eqnarray}
b_{0}&=& (-m^{3}+\frac{\Lambda^{3}}{2}+\frac{m^{2}\Lambda}{2}) ;\ \ \ \ \ \ \  \ \ \ b_{1}= (\frac{m^{4}}{2}-\frac{\Lambda^{4}}{2}) ;\nonumber\\
b_{2}&=&(-\frac{m^{5}}{3!}+\frac{\Lambda^{5}}{2\times 2!}-\frac{m^{2}\Lambda^{3}}{6\times 2!}) ;\ \ 
b_{3}=(\frac{m^{6}}{4!}-\frac{\Lambda^{6}}{2\times 3!}+\frac{m^{2}\Lambda^{4}}{4\times 3!}) ;\nonumber\\
b_{4}&=&(-\frac{m^{7}}{5!}+\frac{\Lambda^{7}}{2\times 4!}-\frac{3 m^{2}\Lambda^{5}}{6\times 4!}) ;\ \ 
b_{5} =(\frac{m^{8}}{6!}-\frac{\Lambda^{8}}{2\times 5!}+\frac{m^{2}\Lambda^{6}}{3\times 5!}) ;\nonumber\\
b_{6} &=&(-\frac{m^{9}}{7!}+\frac{\Lambda^{9}}{2\times 6!}-\frac{5 m^{2}\Lambda^{7}}{14\times 6!}) ;\ \ 
b_{7} = (\frac{m^{10}}{8!}-\frac{\Lambda^{10}}{2\times 7!}+\frac{3 m^{2}\Lambda^{8}}{8\times 7!}) ;\nonumber\\
b_{8} &=& (-\frac{m^{11}}{9!}+\frac{\Lambda^{11}}{2\times 8!}-\frac{7 m^{2}\Lambda^{9}}{18\times 8!}) ;
b_{9} = (\frac{m^{12}}{10!}-\frac{\Lambda^{12}}{2\times 9!}+\frac{2 m^{2}\Lambda^{10}}{5\times 9!}) .
\end{eqnarray}

\section{Solving the Schr\"{o}dinger equation for pentaquark $P_{c}$ }

To investigate the existence of one bound state of pentaquark, we solve the Schr\"{o}dinger equation with calculated potential in previous section for pentaquark $P_{c}$.\\

The radial Schr\"{o}dinger equation for two-body systems is:

  \begin{equation}
  (\frac{d^{2}}{dr^{2}}+\frac{2}{r}\frac{d}{dr}-\frac{l(l+1)}{r^{2}})R_{n,l}(r)+\frac{2\mu }{\hbar}(E-V(r))R_{n,l}(r)=0 
  \end{equation} 
  
 Taking $\hbar=1 $ and changing the variable $\phi(r)= rR_{n,l}(r)$, Eq. (5) becomes:
  
  \begin{equation}
  \frac{d^{2}}{{dr}^{2}}\ \phi(r)+2\mu (E-V(r)-\frac{l(l+1)}{2\mu r^{2}})\phi (r)=0\ \ \ .
  \end{equation} 
  
   By placing the potential $V(r)=V_{\Sigma_{c}\overline{D}^{*}}(r)$ and the expanded shape of $\nabla^{2}Y$ in Eq. (6), an equation is expressed as follows:
  
  \begin{eqnarray}
  &&\frac{d^{2}}{dr^{2}}\phi(r)+2 \mu (E-C_{0}-C_{1}r-C_{2} r^{2}-C_{3} r^{3}-C_{4} r^{4}- C_{5} r^{5}\nonumber\\
  &&-C_{6} r^{6}-C_{7} r^{7}-C_{8} r^{8}-C_{9} r^{9}+...-\frac{l(l+1)}{2 \mu r^{2}})\phi(r)=0.
  \end{eqnarray}
  
  Where:
  
  \begin{eqnarray}
  C_{n}=B_{1} b_{n};\ \ \ \ n=0, 1, ..., 9;\ \ \ \ B_{1}=\frac{1}{12 \pi}\frac{g g_{1}}{f_{\pi}^{2}} \mathcal{J}_{0} \mathcal{G}_{0};\ \ . 
  \end{eqnarray}
   
   In Eq. (8), For $\Sigma_{c}\overline{D}^{*}$  With $I=1/2, J=3/2$, is placing the numerical value of $\mathcal{J}_{0} \mathcal{G}_{0}=1$ product from literature\cite{2}.\\
  
   By considering the following proposed reply (cf. \cite{17, 18}) for the differential Eq. (7) yields:
  
  \begin{equation}
  \phi(r)=N(r) e^{M(r)}= r^{n} e^{M(r)}.
  \end{equation}
  
Differentiating second degree of $\phi$ in Eq.(9) gives:

 \begin{equation}
 \phi''(r)=(n(n-1)r^{-2}+2nM'r^{-1}+M''+{M'}^{2})r^{n} e^{M}.
 \end{equation}
  
  Here, we solving differential equation (7) by considering expansion $V(r)$ up to the $10^{th}$ order for calculating binding energy $E_{B}$ of pentaquark. Hence, an approximation was attempted up to the $10^{th}$ order, not only expanded potential behavior up to the $10^{th}$ order is the same as potential behavior in equation (1), but also resulting of binding energy have sufficient precision compared to numerical literature\cite{2, 19}. This is indicating adequacy of approximation.\\
  
 Now, we can be considered two position for $\phi(r)$. The first $\phi(0)=0$ is for $n=1$ and the second $\phi(0)=cte$ is considered for $n=0$. Thus, we study them in separated subsection. 
 
  \subsection{The position $\phi (0)=0 $ }
  
  To considered $\phi(0)=0 $ then $M(r)$ is Yields\cite{19}:
  
  \begin{eqnarray}
  M(r)&=&a_{1}r^{2}+a_{2}r^{3}+a_{3}r^{4}+a_{4}r^{5}+a_{5}r^{6}+a_{6}r^{7}\nonumber\\
  &&+a_{7}r^{8}+a_{8}r^{9}+a_{9}r^{10}.
  \end{eqnarray}
  
  Replacing Eq. (11) into the Eq. (10) and comparing two Eqs. of (7), (10), the following expression (Eq.(12)) for $r^{-2}$  is obtained in terms of $l(l+1)$ in Eq.(8), as well as, a system of 11 non-linear equations expressed later:
  
  \begin{equation}
  n(n-1)=l(l+1)
  \end{equation}
  
 In the base state, two values for $n$ namely $n=0$ and $n=1$ is obtained that according to Eq. (9) and the condition $\phi (0)=0 $, the value $n=0$ could be unacceptable and $N(r) =r$ will be. Therefore, $\phi (r) $ , $\phi''(r) $ are obtained as follows:

  \begin{eqnarray}
  &&\phi (r)=r e^{M(r)}\nonumber\\
  &&\phi''(r)=(2M'r^{-1}+M''+{M'}^{2})r e^{M}.
  \end{eqnarray}

   After replacing Eq.(11) in Eq. (13) and equal to the different powers of r, Separately, the following nonlinear equations are obtained:
  
  \begin{eqnarray}
  &&6a_{1}=-2 \mu (E-C_{0}) \nonumber\\
  &&12a_{2}=2 \mu C_{1}\nonumber\\
  &&20a_{3}+4a_{1}^{2}=2 \mu C_{2}\nonumber\\
  &&30a_{4}+12a_{1}a_{2}=2 \mu C_{3}\nonumber\\
  &&42a_{5}+9a_{2}^{2}+16a_{1}a_{3}=2 \mu C_{4}\nonumber\\
  &&56a_{6}+20a_{1}a_{4}+24a_{2}a_{3}=2 \mu C_{5}\nonumber\\
  &&72a_{7}+16a_{3}^{2}+24a_{1}a_{5}+30a_{2}a_{4}=2 \mu C_{6}\nonumber\\
  &&90a_{8}+28a_{1}a_{6}+36a_{2}a_{5}+40a_{3}a_{4}=2 \mu C_{7}\nonumber\\
  &&110a_{9}+25a_{4}^{2}+32a_{1}a_{7}+42a_{2}a_{6}+48a_{3}a_{5}=2 \mu C_{8}\nonumber\\
  &&132a_{10}+36a_{1}a_{8}+48a_{2}a_{7}+56a_{3}a_{6}+60a_{4}a_{5}=2 \mu C_{9}.
  \end{eqnarray} 
   
  Here, by replacing the numerical values of constants for pentaquark $P_{c}(4380)$, $\Sigma_{c}=2455 MeV$ and $\overline{D}^{*}=2008.32 MeV$, we obtained the binding energy of pentaquark\cite{2}. In table 1, 2 have been shown different values of binding energy from Eq. (14) for $P_{c}(4380)$.\\
 
  \begin{table}[ht]
  	\begin{center}
  		\caption{ $(E_B (MeV) , M_{P_{c}} (MeV))$ for  $P_{c}(4380)$  and $g_{1}=0.94$  }
  		\onehalfspacing
  		\begin{tabular}{c|c|c|c}
  			\hline{$\Lambda (MeV)$} & {$g=0.51$} & {$g=0.59$} & {$g=0.67$}  \\
  			\hline\hline{ $800 $} & {$(-63, \ 4400.32)$} & {$(-40, \ 4423.32)$} & {$(-17, \ 4446.32) $}  \\
  			{ $850 $} & {$(-62, \ 4401.32)$}& {$(-34, \ 4429.32)$}& {$(-6.7, \ 4456.62)$} \\ 
  			{ $900 $} & {$(-59, \ 4404.32)$}& {$(-26, \ 4437.32)$}& {-} \\ 
  			{ $1000 $} & {$(-45, \ 4418.32)$}& {$(-0.11, \ 4463.21)$}& {-} \\ 
  			{ $1100 $} & {$(-22, \  4441.32)$}& {-}& {-} \\ 
  			\hline
  		\end{tabular}
  	\end{center}
  \end{table}

\begin{table}[ht]
	\begin{center}
		\caption{ $(E_B (MeV) , M_{P_{c}} (MeV))$ for $P_{c}(4380)$  and $g_{1}=0.75$  }
		\onehalfspacing
		\begin{tabular}{c|c|c|c}
			\hline{$\Lambda (MeV)$} & {$g=0.51$} & {$g=0.59$} & {$g=0.67$}  \\
			\hline\hline{ $800 $} & {$(-90, \ 4373.32)$} & {$(-73, \ 4390.32)$} & {$(-56, \ 4407.32)$}  \\
			{ $850 $} & {$(-95, \ 4368.32)$}& {$(-75, \ 4390.32)$}& {$(-53, \ 4407.32)$} \\ 
			{ $900 $} & {$(-99, \ 4364.32)$}& {$(-74, \ 4389.32)$}& {$(-49, \ 4414.32)$} \\ 
			{ $1000 $} & {$(-102, \ 4361.32)$}& {$(-67, \ 4369.32)$}& {$(-32, \ 4431.32)$} \\ 
			{ $1100 $} & {$(-98, \ 4365.32)$}& {$(-51, \ 4412.32)$}& {$(-3.14, \ 4460.18)$} \\ 
			{ $1200 $} & {$(-86, \ 4377.32)$}& {-}& {-} \\ 
			\hline
		\end{tabular}
	\end{center}
\end{table}

In the tables above $M_{P_{c}}$ is calculated as follows:

\begin{eqnarray}
M_{P_{c}}=M_{\Sigma_{c}}+M_{\overline{D}^{*}}+E_{B}.
\end{eqnarray} 

According to the obtained values, it is observed that in $g_{1}=0.75$, the results for pentaquark mass are much closer to the $M_{P_{c}(4380)}$. Also, to check the results, one of the main differences between this paper and the other references \cite{2, 20} is that the acceptable results are obtained for the $M_{P_{c}(4380)}$ only in the $800 MeV \leq \Lambda \leq 1200 MeV$ and these can't be found in $\Lambda \geq 1200 MeV$.

  \subsection{The position $\phi (0)=cte $ }
  
   Now, we considered $\phi (0)=cte $, thus $M(r)$ will be as follows\cite{19}:
  
  \begin{eqnarray}
  M(r)=&&a_{1}r+a_{2}r^{2}+a_{3}r^{3}+a_{4}r^{4}+a_{5}r^{5}+a_{6}r^{6}\nonumber\\
  &&+a_{7}r^{7}+a_{8}r^{8}+a_{9}r^{9}.
  \end{eqnarray}
  
  Here, replacing Eq. (15) into the Eq. (10) and comparing to Eq. (7), a similar equation is obtained by Eq. (12). In this position, i.e. $\phi (0)=cte$, the value $n=0$ could be acceptable and $N(r) =1$ will be. Therefore, $\phi (r)$ , $\phi''(r)$ are obtained as follows: 
  
  \begin{eqnarray}
  &&\phi (r)= e^{M(r)}\nonumber\\
  &&\phi''(r)=(M''+{M'}^{2}) e^{M}.
  \end{eqnarray}
  
 The same as before, by replacing Eq.(15) in Eq. (16), 10 nonlinear equations are obtained, we have:
 
  \begin{eqnarray}
  &&2a_{2}+a_{1}^{2}=2 \mu (C_{0}-E) \nonumber\\
  &&6a_{3}+4a_{1}a_{2}=2 \mu C_{1}\nonumber\\
  &&12a_{4}+4a_{2}^{2}+6a_{1}a_{3}=2 \mu C_{2}\nonumber\\
  &&20a_{5}+8a_{1}a_{4}+12a_{2}a_{3}=2 \mu C_{3}\nonumber\\
  &&30a_{6}+9a_{3}^{2}+10a_{1}a_{5}+16a_{2}a_{4}=2 \mu C_{4}\nonumber\\
  &&42a_{7}+12a_{1}a_{6}+20a_{2}a_{5}+24a_{3}a_{4}=2 \mu C_{5}\nonumber\\
  &&56a_{8}+16a_{4}^{2}+14a_{1}a_{7}+24a_{2}a_{6}+30a_{3}a_{5}=2 \mu C_{6}\nonumber\\
  &&72a_{9}+16a_{1}a_{8}+28a_{2}a_{7}+36a_{3}a_{6}+40a_{4}a_{5}=2 \mu C_{7}\nonumber\\
  &&25a_{5}^{2}+18a_{1}a_{9}+32a_{2}a_{8}+42a_{3}a_{7}+48a_{4}a_{6}=2 \mu C_{8}\nonumber\\
  &&36a_{2}a_{9}+48a_{3}a_{8}+56a_{4}a_{7}+60a_{5}a_{6}=2 \mu C_{9}.
  \end{eqnarray} 
 
 Also, in this position for $P_{c}(4380)$, we are obtained the binding energy of pentaquark and the numerical coefficients of wave function. In table 3, has been shown the amount of binding energy from Eq. (17) and $\Lambda=800 MeV$ for $P_{c}(4380)$.\\
 
\begin{table}
	\begin{center}
		\caption{ $(E_B (MeV) , M_{P_{c}} (MeV))$  for $P_{c}(4380)$  and $\Lambda=800 MeV$ }
		\onehalfspacing
		\begin{tabular}{c|c|c|c}
			\hline{$g_{1}$} & {$g=0.51$} & {$g=0.59$} & {$g=0.67$}  \\
			\hline\hline{ $0.75$} & {$(-15.99, \ 4447.33)$} & {$(-15.80, \ 4447.52)$} & {$(-15.62, \ 4447.7)$}  \\
			{ $0.94$} & {-}& {$(-15.47, \ 4447.85)$}& {$(-15.27, \ 4448.05)$} \\ 
			{ $1.95$} & {-}& {$(-14.01, \ 4449.32)$}& {-} \\ 
			\hline
		\end{tabular}
	\end{center}
\end{table}

To confirm the existence of pentaquark states, binding energy must be negative i.e. $E_{B}<0$.  Also, the total mass of particles contributing of pentaquark(i.e. the sum of baryon $\Sigma_{c}$ and meson $\overline{D}^{*}$ masses) in addition to binding energy is closer to the mass of pentaquark $P_{c}(4380)$. Here, the obtained results indicate binding energy ranging $-102 \leq E_{B} \leq -0.11$ for pentaquark $P_{c}(4380)$ which are negative. Also, that conform to calculated results in the literatures\cite{2, 20, 21, 22, 23}. According to mentioned above, obtained results are acceptable to a great extent and they could be considered a clear evidence for the existence of a bound five-quark state.\\

Figure 2, demonstrates the wave function's diagrams for pentaquark $P_{c}$ in $\Lambda=800 MeV$, at different values of $g$ and $g_{1}$. These charts tend to zero at the given value. As shown in the graphs, wave functions become zero in $x \sim 0.12-0.13$, indicating that the maximum pentaquark radius is ranges from 23.67 to 25.64 fm.

\begin{center}
	\begin{figure}
		
		\includegraphics[width=8 cm, totalheight=6 cm]{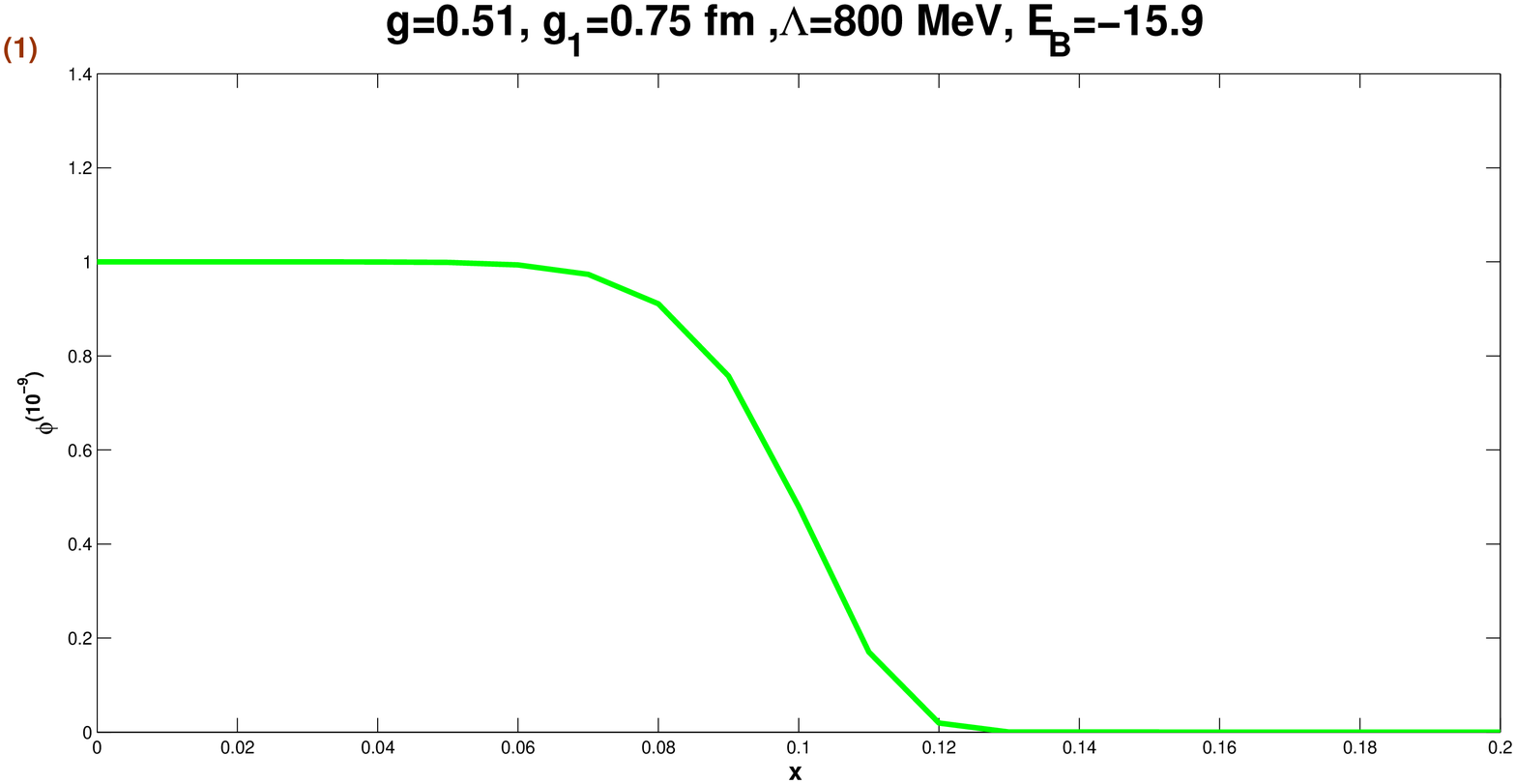}
		\includegraphics[width=8 cm, totalheight=6 cm]{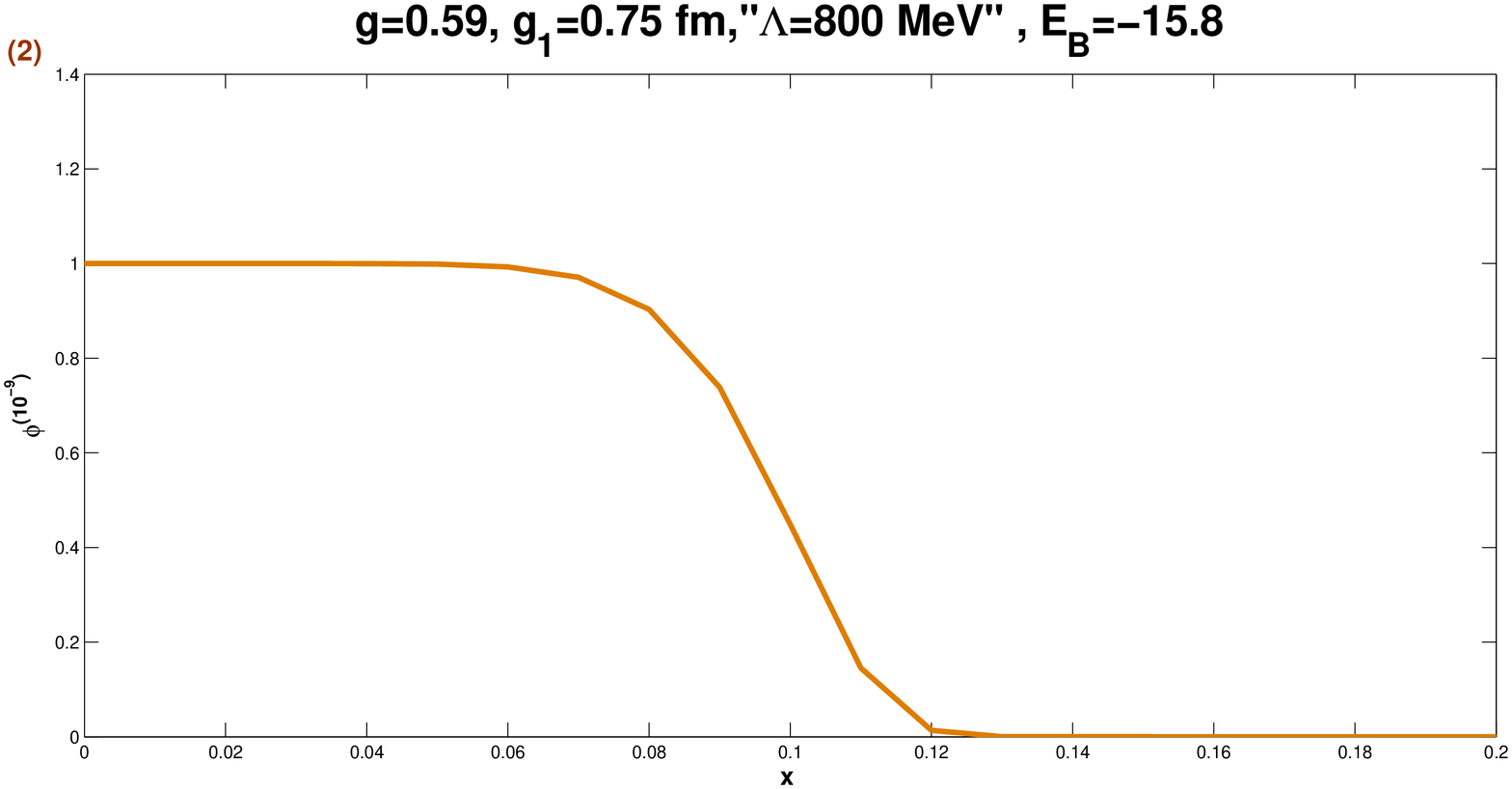}
		\includegraphics[width=8 cm, totalheight=6 cm]{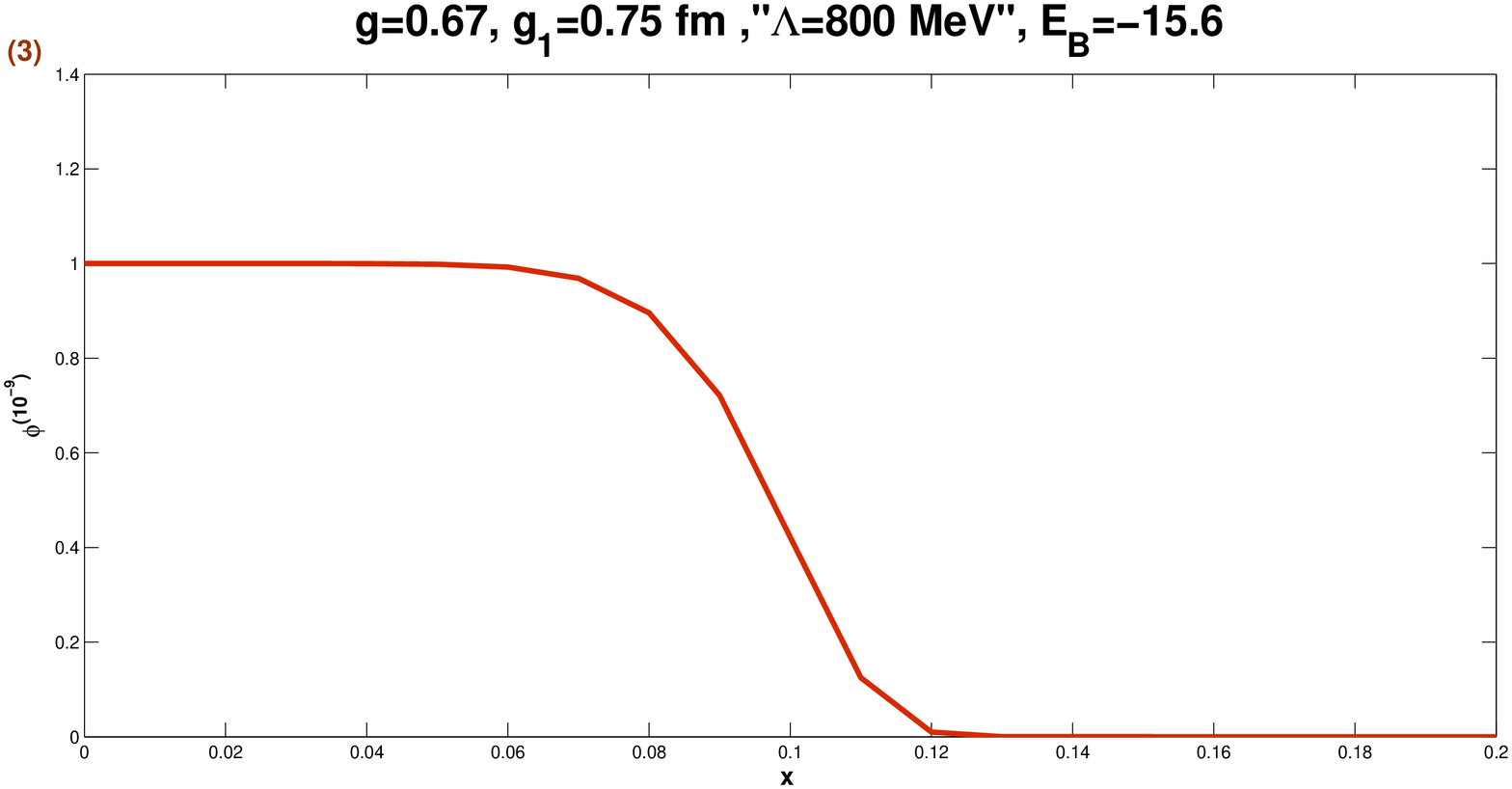}
		\includegraphics[width=8 cm, totalheight=6 cm]{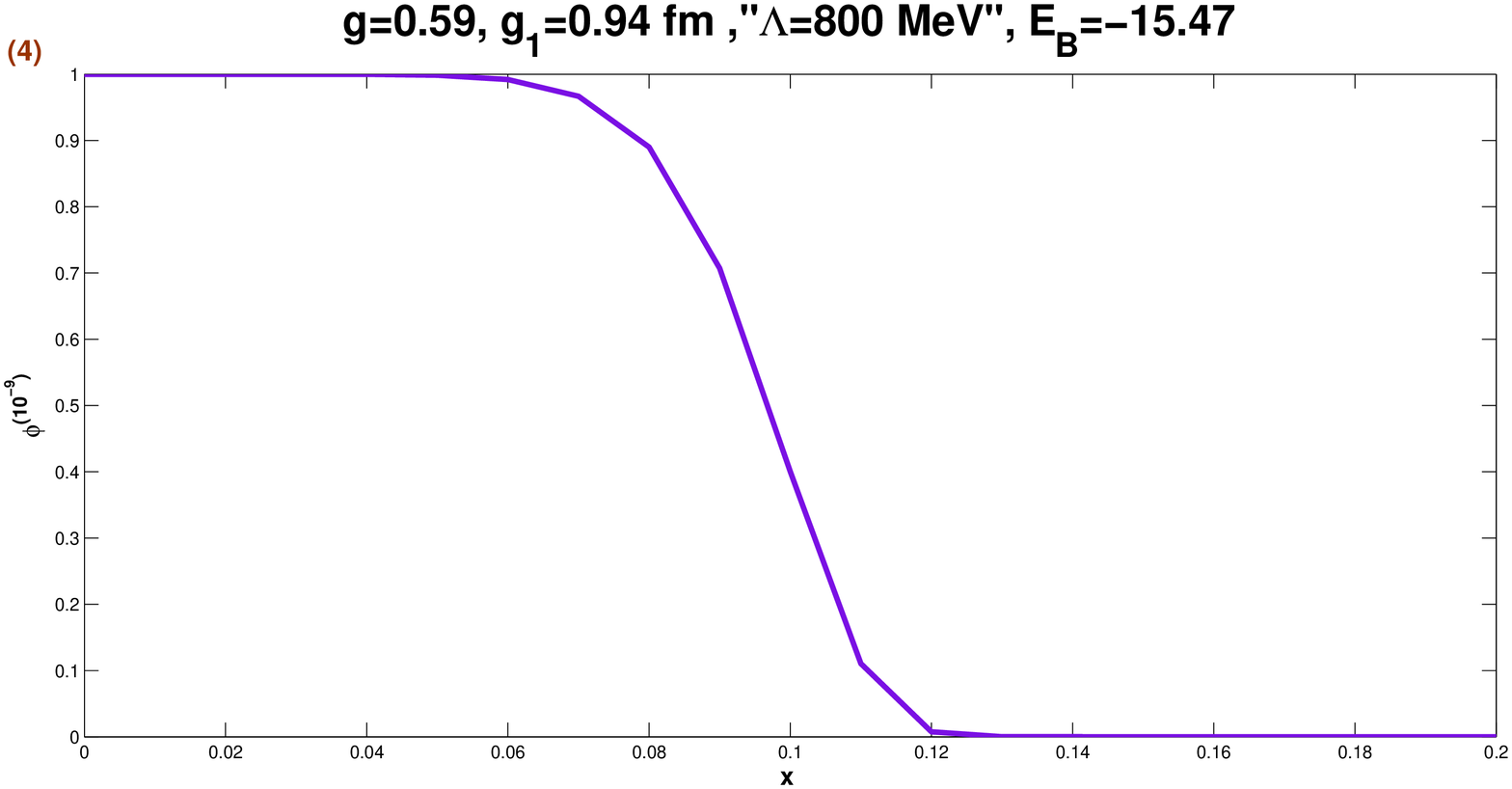}
		\includegraphics[width=8 cm, totalheight=6 cm]{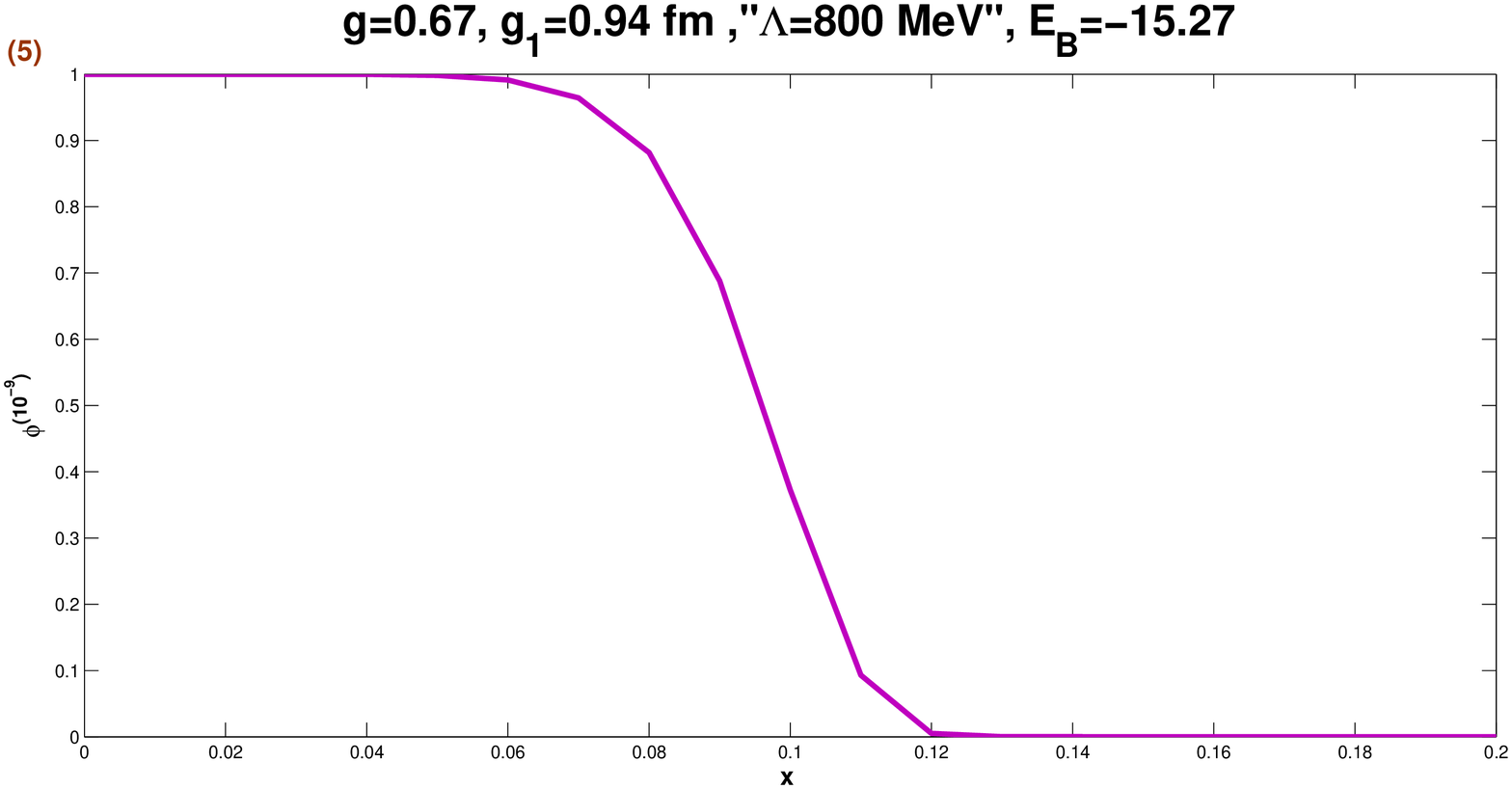}
        \includegraphics[width=8 cm, totalheight=6 cm]{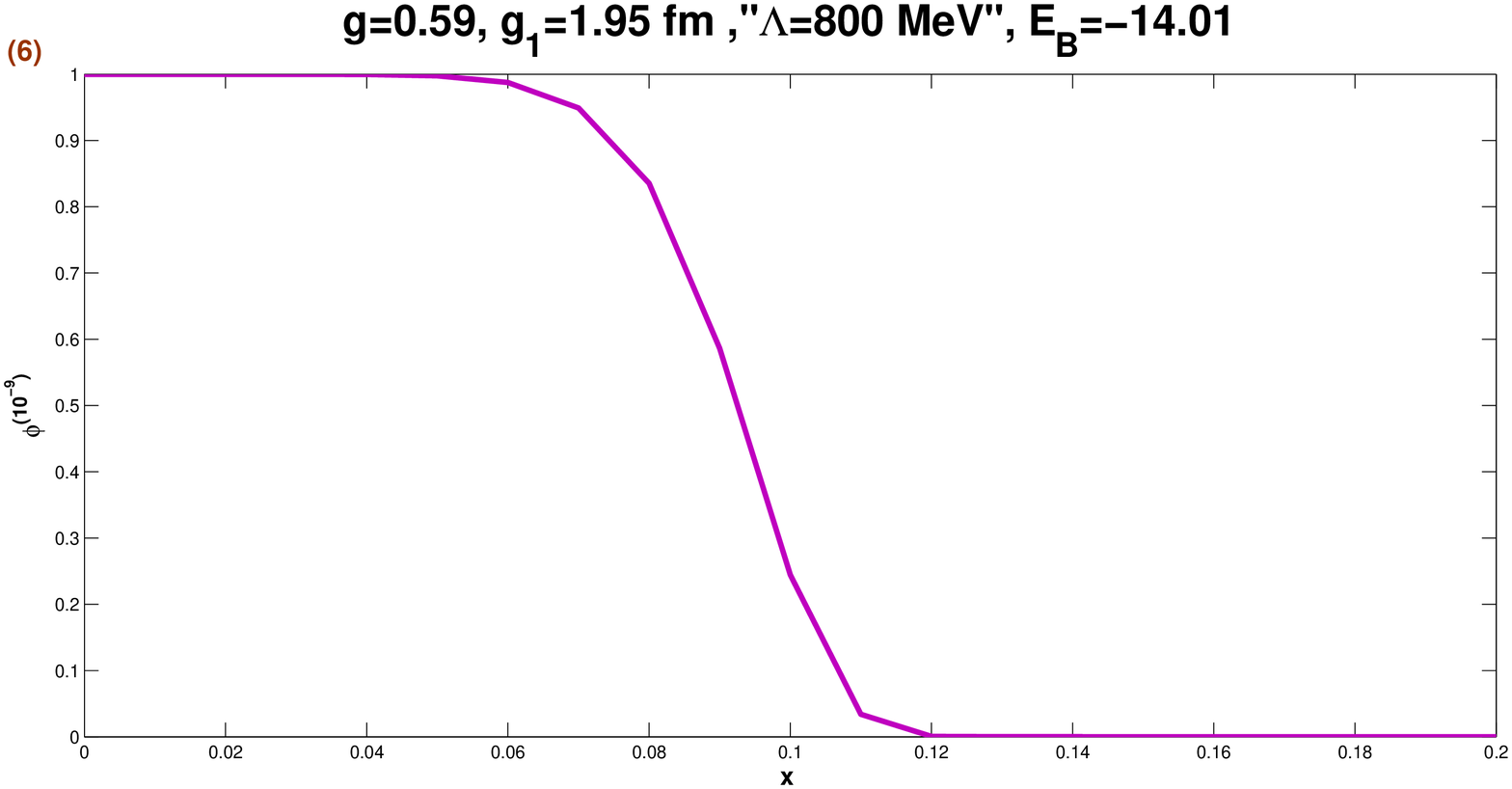}
		\caption{\it\small{{$P_{c}(4380)$ at $r$ in $\Lambda=800 MeV$ for different values of $g$ and $g_{1}$ }}}
		\label{fig:2}
	\end{figure}
\end{center}

\section{CONCLUSIONS}
 
  In this article, pentaquark $P_{c}(4380)$ system consisting of baryon $\Sigma_{c}$ and $\overline{D}^{*}$ meson has been considered. The obtained potential for pentaquark in reference\cite{2} was presented and expanded. Then, expanded potential was replaced in the Schr\"{o}dinger equation and that was solved as a bound state of two-body systems. By solving this to analytically approach and according to the values of constants and cutoff, 10 nonlinear differential equations and binding energy $E_{B}$ of pentaquark $P_{c}$ and wave function  coefficients were obtained. Results $E_{B}$ and wave function were presented as tables and diagrams in the previous section, which could be confirmed the existence of a bound state of pentaquark $P_{c}(4380)$. Then, it specified that the wave function plots tend to be zero at a given value. Therefore, the maximum radius of pentaquark $P_{c}$ were found out which ranged from $x=23.67 fm$ to $x=25.64 fm$. We observed that the calculated values are matches good with others and the mass of $P_{c}(4380)$. Also, we can say that advantages of this study are that results obtained analytically while other references are numerically calculated. Hence, the results are more comprehensive and complete than other articles.
  
  \section{Data Availability}
  The authors confirm that the data supporting the findings of this study are available within the article and its supplementary materials.
  
 \section{Disclosure}
  The research was performed as part of the employment of the authors.
  
  \section{Conflicts of Interest}
  The authors also declare that there is no conflict of interests regarding the publication of this paper.

\end{document}